\documentclass[twocolumn,showpacs,superscriptaddress,amsmath,amssymb,nofootinbib]{revtex4}
\usepackage{graphicx,psfrag}
\usepackage{dcolumn}
\usepackage{bm}
\usepackage{subfigure}
\usepackage{amsmath}
\usepackage{amssymb}

\begin{document}

\title{Formation of $\eta$-mesic nuclei by ($\pi,N$) reaction
and $N^*(1535)$ in medium}

\author{Hideko Nagahiro}
   \affiliation{Research Center for Nuclear Physics(RCNP), Osaka
University, Ibaraki, Osaka, 567-0047, Japan}

\author{Daisuke Jido}
   \affiliation{Yukawa Institute for Theoretical Physics, 
Kyoto University, Kyoto 606-8502, Japan}

\author{Satoru Hirenzaki}
   \affiliation{Department of Physics, Nara Women's University, Nara, 630-8506, Japan}
\begin{abstract}

We calculate formation spectra of $\eta$-nucleus systems in 
($\pi,N$) reactions with nuclear targets, which can be performed 
at existing and/or forthcoming facilities, including J-PARC, 
in order to investigate $\eta$-nucleus interactions. 
Based on  the $N^*(1535)$ dominance in the $\eta N$ system,
$\eta$-mesic nuclei are suitable systems for study of 
in-medium properties of the $N^*(1535)$ baryon resonance,
such as reduction of the mass difference of $N$ and $N^{*}$
in nuclear medium, which affects level structure of the $\eta$ and 
$N^{*}$-{\it hole} modes. 
We find that clear information on the in-medium $N^*$- and 
$\eta$-nucleus interactions can be obtained through the formation spectra 
of the $\eta$-mesic nuclei.
We also discuss the experimental feasibilities by showing several spectra
of ($\pi,N$) reactions calculated with possible experimental settings.
Coincident measurements of $\pi N$ pairs from the $N^{*}$ decays 
in nuclei help us to reduce backgrounds.



\end{abstract}
\maketitle

\section{Introduction}
\label{intro}
The study of meson-nucleus bound systems is
{one of the important subjects}
in nuclear physics.
The detailed investigations
{of structure of bound states}
provide us quantitative information on hadron-nucleus
interactions.  So far, the structure of atomic states of pion, kaon, and $\bar{p}$ 
have been successfully observed and investigated comprehensively 
both in theoretical and experimental points of view~\cite{Batty97}. 
One of the remarkable developments in experimental aspects is the
establishment of (d,$^{3}$He) spectroscopy for the formation of
deeply bound pionic atoms with recoil free kinematics~\cite{NPA530etc,PRC62,PRL88,PLB514}.
It opens new possibilities of the formation of other hadron-nucleus
bound systems~\cite{EPJA6,PLB443etc,NPA710,PRC66(02)045202,PRC68(03)035205}.

The bound states of the $\eta$ meson in nuclei were predicted first by 
Haider and Liu~\cite{Haider}. After that many works were devoted 
to studies of the structure of the bound states, the formation reactions
of $\eta$-mesic nuclei and in-medium properties of the $\eta$ meson~\cite{Chiang:1990ft,PLB231,Kohno:1990xv,Waas:1997pe,EPJA6,PLB443etc,Inoue:2002xw,PRC66(02)045202,PLB550,PRC68(03)035205,Nagahiro:2005gf,Kelkar:2006zs,levelcross,Song:2008ss}.
Especially,  the $\eta$ meson in nuclear medium 
has been recently investigated in the aspect of chiral symmetry~\cite{Waas:1997pe,EPJA6,PLB443etc,Inoue:2002xw,PRC66(02)045202,PLB550,PRC68(03)035205,Nagahiro:2005gf,levelcross}.
The $\eta$-nucleus systems are purely governed
by strong interactions in contrast to the atomic states of mesons with negative charge.
Thus, the $\eta$ mesons in the bound states are largely overlapped 
with nuclei. In such compact systems, large medium effects on the mesons inside nuclei 
are expected, and, at the same time, wide natural widths of the bound states due to absorptions of the mesons into the nucleus
are inevitable, as seen deeply bound kaonic nuclei~\cite{Yamagata:2006sm}.


The first experimental search of the $\eta$ bound states in nuclei~\cite{Chrien:1988gn}
was performed in $(\pi^{+},p)$ reactions with several nuclear targets 
in finite momentum transfer to aim to observe narrow states as predicted 
in Ref.~\cite{Haider}, and the result turned out to be negative. 
Some hints of the $\eta$ bound states were also observed as
enhancement at the subthreshold of the $\eta$ meson production
in d(p,$^{3}$He)$\eta$~\cite{Berger:1988ba} and
$^{18}$O($\pi^{+}$,$\pi^{-}$)$^{18}$Ne~\cite{Johnson:1993zy}
reactions. Observation of a $\eta$ 
meson bound state in $^{3}$He was reported in photoproduction
reactions~\cite{Pfeiffer:2003zd}, though interpretation of  these
observations are still controversial~\cite{Hanhart:2004qs}. 
It was suggested in Ref.~\cite{sokol} that coincident observation 
of $\pi N$ pairs from $N^{*}(1535)$ helps to identify the formation of $\eta$ 
meson bound states. Other experiments have been also proposed~\cite{Baskov:2003vr,Anikina:2004ps,Kenta,Jha}.

The study of the hadron properties in nuclear medium is
largely related to the fate of chiral symmetry in finite density.
It is expected that partial restoration of chiral symmetry
in nuclear medium takes place as reduction of the quark condensates~\cite{PR247etc,Jido:2008bk} and provides effective change
of the hadron properties. 
For the context of the study of the $\eta$ meson in nuclear medium,
the $N^{*}(1535)$ resonance, which can be a candidate 
of the chiral partner of the nucleon becoming degenerate in the chiral 
restoration limit~\cite{DeTar:1988kn,Jido:1998av,Jido:2001nt},
plays an important role in the $\eta$-mesic nuclei
due to the strong coupling of the $\eta$-nucleon system 
to the $N^{*}(1535)$ resonance.
In our previous works~\cite{PRC66(02)045202,PRC68(03)035205,Nagahiro:2005gf},
it was found that the $\eta$ optical potential in nuclei is strongly sensitive to the
in-medium mass gap of $N$ and $N^{*}(1535)$ and that, as a consequence,
the formation spectra of the $\eta$-mesic nuclei is also sensitive to the 
in-medium properties of the $N^{*}(1535)$. 

The sensitivity of the $\eta$-nucleus optical potential to the $N$-$N^{*}$ mass 
gap stems from possible level crossing  
between $N^*$-{\it hole} and $\eta$ modes in  nuclear medium as suggested 
in Ref.~\cite{levelcross}.
The level difference between the $N^{*}$-{\it hole} and $\eta$
modes without medium effects is only 50 MeV, which is 10\% of 
the $N$-$N^{*}$ mass gap and small enough in energy scale 
of hadronic interactions. We found that the level crossing 
caused by the reduction of the mass gap provides deep $\eta$ bound states 
and significant enhancement 
of the formation spectra of the $\eta$-mesic nuclei 
in the quasi-free $\eta$ production energies.
This will be a clue to deduce the in-medium $N^{*}$ properties 
for the $\eta$-mesic nuclei. 


In this paper, 
we revisit 
the ($\pi,N$) reaction with the recoil free kinematics for  
the formation of $\eta$-mesic nuclei 
in order to get clearer information on the level structure of
the $\eta$ and $N^{*}$-{\it hole} modes and
in-medium properties of $N^*(1535)$ in
the viewpoint of the chiral symmetry for baryons. 
We will find that the appropriate kinetic energy of the injecting pion
in this reaction can be attained by the Japan Proton Accelerator
Research Complex (J-PARC) facility. We will also compare our 
calculation with the old experiments of the ($\pi^+$,p) reaction 
with finite momentum transfer~\cite{Chrien:1988gn}, in which
the expected peak structures predicted in Ref.~\cite{Haider}
were not found.
In this paper, we discuss more appropriate experimental conditions 
than the old experiments and also propose the coincidence observation
to reduce large background, which would be one of the reasons
why the experiment~\cite{Chrien:1988gn} could not see the
expected peak structure.

This paper is organized as follows. In Sec.~\ref{sec:level_cross}, we
discuss the properties of the 
$\eta$ spectral functions in the nuclear matter. 
In Sec.~\ref{sec:potential}, we
introduce the $\eta$-nucleus optical potentials with a finite size
nucleus and discuss their features.
The formation spectra of $\eta$-mesic nuclei will be
shown in Sec.~\ref{sec:formulation},
and the physical meaning of the formation
spectra will be discussed.
In Sec.~\ref{sec:feasability} we will
discuss the experimental feasibilities, and 
finally, we will devote Sec.~\ref{sec:conclusion} to summary of this
paper.

\section{Level crossing of the $\eta$ meson and $N^*$-hole modes}
\label{sec:level_cross}
In this section, we briefly review the interesting feature of the $\eta$
meson in nuclear medium pointed out in Ref.~\cite{levelcross}.
Reduction of the mass gap between $N$ and $N^{*}$ causes 
level crossing of the $\eta$ and $N^{*}$-{\it hole} modes, and 
consequently the $\eta$ spectral function in nuclear medium 
has significant properties. 

The in-medium $\eta$ propagator is given by,
\begin{equation}
D_\eta(\omega,k;\rho)^{-1}=\omega^2-k^2-m_\eta^2-\Pi_\eta(\omega,k;\rho),
\label{eq:G_eta}
\end{equation}
where $\omega$ and $k$ denote the energy and momentum of the $\eta$
meson, $m_\eta$ is its mass, and $\Pi_\eta$ denotes the $\eta$
self-energy in the nuclear medium.
Because of the strong coupling of the $\eta N$ system to the $N^*(1535)$
resonance, the $\eta$ self-energy $\Pi_\eta$ can be evaluated by the $N^*$
dominance. 
Considering one $N^{*}$-nucleon-hole excitation, we obtain the $\eta$ self-energy
in small $\eta$ momentum~\cite{PRC66(02)045202,levelcross} as 
\begin{eqnarray}
\Pi_\eta(\omega,k;\rho)&=&
\frac{g_\eta^2\rho}{\omega+m^*_N(\rho)-m^*_{N^*}(\rho)+i\Gamma_{N^*}(\omega,\rho)/2}\nonumber\\
&+& {\rm (cross\ term)}.
\label{eq:Pi_eta}
\end{eqnarray}
Here, $g_\eta$ is the coupling constant of the $s$-wave $\eta NN^*$ vertex
and can be determined to $g_\eta\simeq 2.0$ to reproduce the in-vacuum partial width
$\Gamma_{N^*\rightarrow \eta N} \simeq 75$ MeV~\cite{Amsler:2008zz} at tree
level. 
$m_N^*(\rho)$ and $m_{N^*}^*(\rho)$ are effective masses
of $N$ and $N^*$ in the 
nuclear medium with density $\rho$, respectively. 
\begin{figure}[hbt]
\center
\resizebox{0.48\textwidth}{!}{
\includegraphics{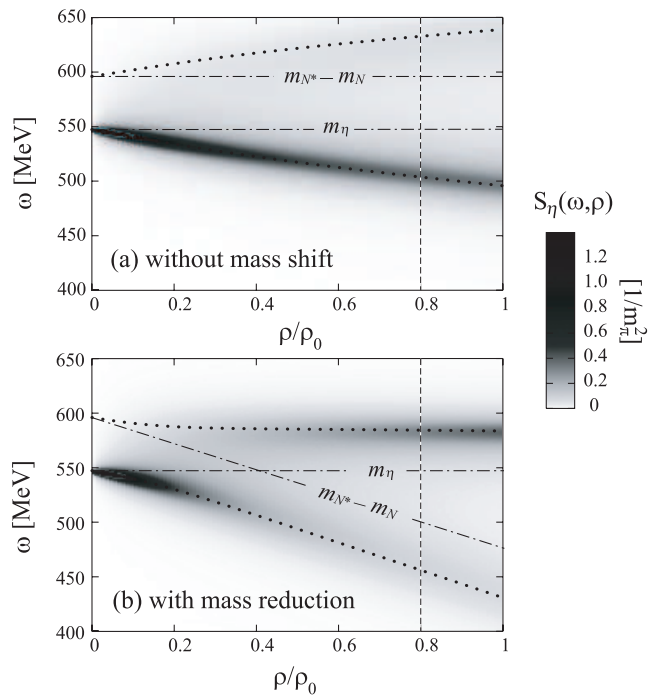}}
\caption{Contour maps of the $\eta$ meson spectral density in nuclear
 matter in Eq.~(\ref{eq:S_eta})
as functions of the baryon density and energy
assuming (a) the $N$ and $N^*$ masses not to change in medium and
 (b) 20\% mass gap reduction of $N$ and $N^*$ at normal nuclear density
 $\rho_0$. 
 In this figure, the $N^*$ width in medium is fixed to be constant
 $\Gamma_{N^*}=75$ MeV for simplisity.
The dotted lines indicate the real parts of the solutions of
 $D_\eta(\omega,k=0;\rho)^{-1}=0$ in Eq.~(\ref{eq:G_eta}).
}
\label{fig:map}       
\end{figure}
\begin{figure}[hb]
\centering
\resizebox{0.48\textwidth}{!}{%
\includegraphics{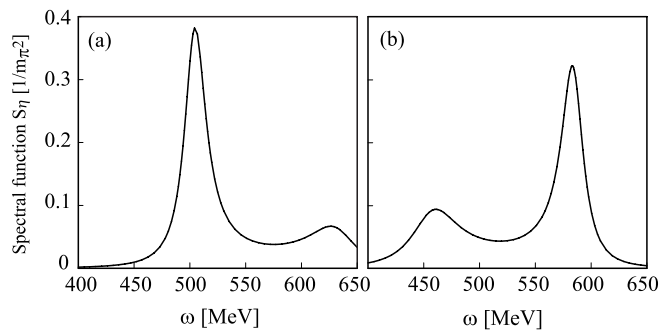}}
\caption{Spectral functions of the $\eta$ meson as functions of
the $\eta$ energy at $\rho/\rho_0=0.8$ (indicated by the vertical dashed lines in
 Fig.~\ref{fig:map}) (a) without mass shift and (b) with $20\%$ mass
gap reduction of $N^*$ and $N$ at normal nuclear density.}
\label{fig:spec}
\end{figure}

The $\eta$ propagator (\ref{eq:G_eta}) with the self-energy (\ref{eq:Pi_eta}) 
has two poles with a positive real part in the complex energy plane in each density. 
These poles describe the $\eta$ meson and $N^*$-{\it hole} modes 
in nuclear medium~\cite{Waas:1997pe,Inoue:2002xw,levelcross}.
Corresponding to these poles,  
the $\eta$ spectral density $S_\eta$ given by
\begin{equation}
S_\eta(\omega,\rho)=-\frac{1}{\pi}{\rm Im}(D_\eta(\omega,k=0;\rho))
\label{eq:S_eta}
\end{equation}
has two peaks in a function of real energy at a certain density.

We show
the contour maps of the $\eta$ spectral density
as functions of baryon density and $\eta$ energy in Fig.~\ref{fig:map},
where the real parts of the pole positions are indicated by dotted lines.
Figure~\ref{fig:map}(a) shows the strength of two branches in the case
that the effective masses of $N$ and $N^*$ do not change in medium.
In this case, two
branches slightly come away from each other for higher $\rho$ as a
result of level repulsion,  
and the strength of the lower mode is always larger than the upper mode
as also shown explicitly in Fig.~\ref{fig:spec}(a).
The similar behavior of the $\eta$ spectral function based on the chiral
unitary approach were also reported
in Refs.~\cite{Waas:1997pe,Inoue:2002xw},
where the reduction of the mass gap between $N$ and $N^*$ is very small.
In contrast, in the case that the mass gap becomes smaller in nuclear medium, 
the behavior of the $\eta$ spectral density significantly
changes.  
Suppose that the mass gap of $N$ and $N^*$ linearly decreases by $20\%$ at $\rho_0$,
the level crossing between two branches takes place around
$\rho\sim 
0.4\rho_0$ as shown in 
Fig.~\ref{fig:map}(b).
As a consequence, the
strength of the upper mode 
becomes stronger due to the 
level mixing, and the lower mode shifts
downwards considerably as the density increases. 

A possible source of the mass gap reduction is the partial restoration 
of chiral symmetry in nuclear medium. If $N^{*}(1535)$ is a chiral 
partner of nucleon, the $N$ and $N^{*}$ mass difference should 
decrease as chiral symmetry is being restored.   
For the later discussion, we use the following 
parameterization of the mass gap reduction based on the chiral doublet model~\cite{Jido:2001nt,Jido:1998av}:
\begin{equation}
m^*_{N^*}(\rho) - m^*_{N}(\rho) = \left(1-C\frac{\rho}{\rho_0}\right)(m_{N^*}-m_{N}), 
\label{eq:massdif}
\end{equation}
where $m_N$ and $m_{N^*}$ are the $N$ and $N^*$ masses in free space,
respectively. 
Here the parameter $C$ represents the strength of the chiral restoration
at the normal nucleon saturation density $\rho_0$, and its empirical
value lies from $0.1$ to $0.3$~\cite{Hatsuda:1999kd}. 
Figures \ref{fig:map}(b) and \ref{fig:spec}(b) correspond to the case
with $C=0.2$~\cite{Kim:1998up} in the chiral doublet model.

These characteristic phenomena caused by the level crossing can be 
a signal of the reduction of the $N$ and $N^{*}$ mass gap, which 
supports the partial restoration of the chiral symmetry 
in nuclear medium.
In next section, we introduce the $\eta$-nucleus optical potentials for the study 
in finite size systems.

\section{$\eta$-nucleus optical potentials}
\label{sec:potential}
\begin{figure*}[bth]
\center
\resizebox{1\textwidth}{!}{%
\includegraphics{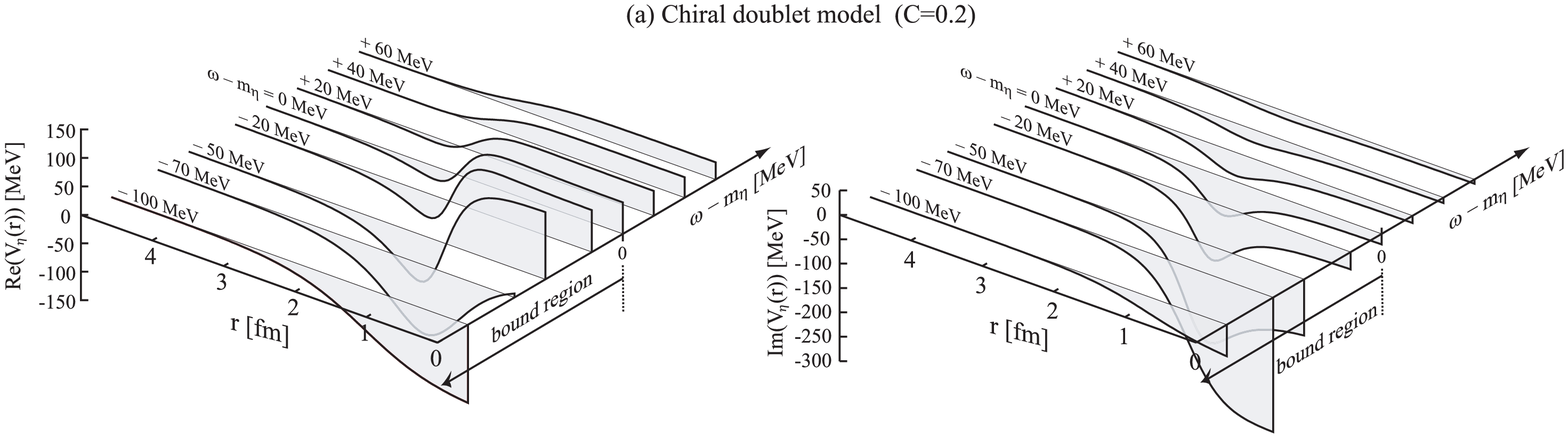}}
\resizebox{1\textwidth}{!}{%
\includegraphics{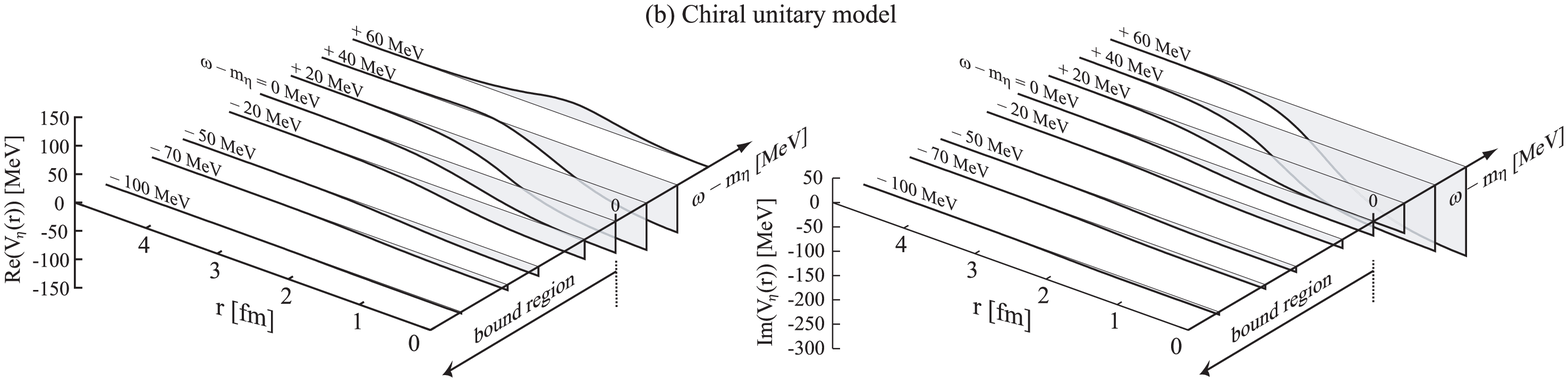}}
\caption{The $\eta$-nucleus optical potentials with (a) the chiral
 doublet model ($C=0.2$) and (b) the chiral unitary model as functions
 of the radial coordinate $r$
for $\eta$ energies $\omega-m_\eta=-100$, $-70$, $-50$, $-20$, $0$,
 $+20$, $+40$, $+60$ MeV.
Left and right figures show the real and imaginary parts of the optical
 potentials $V_\eta$, respectively.
}
\label{fig:potential_3d}       
\end{figure*}

As discussed in the previous section, with the sufficient reduction 
of the $N$-$N^{*}$ mass gap, the level crossing between the $\eta$
and $N^{*}$-{\it hole} modes takes place at certain density 
in nuclear medium. As a consequence of the level crossing, the in-medium
$\eta$ self-energy has strong energy dependence. In addition,
the mass gap reduction as density increases gives also strong density 
dependence on the $\eta$ self-energy as pointed out 
in Ref.~\cite{PRC66(02)045202}. In this section we see that
the $\eta$ optical potential in nuclei also has these features, and 
we briefly explain
the models for the in-medium $N^{*}$ which we use in the calculation
of the formation spectra of the $\eta$-mesic nuclei. 
The details are described in Refs.~\cite{PRC66(02)045202,PRC68(03)035205,Nagahiro:2005gf,levelcross}.


For the in-medium properties of $N^*$, we use 
two kinds of the chiral models, which are based on
distinct physical pictures of $N^*$. 
One is the chiral doublet model~\cite{DeTar:1988kn,Jido:2001nt,Jido:1998av},
in which $N^*$ is regarded as the
chiral partner of the
nucleon.
The 
other is the chiral unitary model, in which $N^*$ is dynamically
generated
resonance
in the coupled channel meson-baryon scattering~\cite{PLB550,Inoue:2002xw}.

In the first approach, the $N^*$ is introduced as a particle with a
large width and appears in an effective Lagrangian together with the
nucleon field in linear realization of chiral symmetry. 
The $\eta$-nucleus optical potential can be obtained from the $\eta$
self-energy~(\ref{eq:Pi_eta}):
\begin{multline}
V_\eta(\omega,r)\equiv\frac{\Pi_\eta(\omega,k=0;\rho(r))}{2\mu}\\
=\frac{g_\eta^2}{2\mu}
\frac{\rho(r)}{\omega+m^*_N(\rho(r))-m^*_{N^*}(\rho(r))+i\Gamma_{N^*}(\omega,\rho(r))/2}\\
+ {\rm (cross\ term)}.
\label{eq:potential}
\end{multline}
Here we use
the local density approximation and heavy baryon
limit \cite{Chiang:1990ft}, and the density-dependent  mass difference 
$m^*_N - m^*_{N^*}$ are given in Eq.~(\ref{eq:massdif}).
We can ignore the momentum $k$ of the $\eta$
meson because we
consider almost recoilless production of the $\eta$ meson in the following
sections.
We use an empirical density distribution of nucleons in 
Woods-Saxon form:
\begin{equation}
\rho(r)=\frac{\rho_0}{1+\exp\left(\frac{r-R}{a}\right)},
\end{equation}
with $R=1.18A^{1/3}-0.48$ fm, $a=0.5$ fm and the nuclear mass
number $A$.


The optical potential~(\ref{eq:potential}) is sensitive to the mass
difference of $N$ and $N^*$ in nuclei. Especially, 
the sign of the real part of the $\eta$ optical potential changes 
when the mass gap of $N^*$ and $N$ becomes smaller than the $\eta$ energy
$\omega$~\cite{PRC66(02)045202,PRC68(03)035205,Nagahiro:2005gf}.
This means that the attractive $\eta$-nucleus interaction at low densities 
can turn to be repulsive depending on the values of the mass gap and
the $\eta$ energy. This feature can be seen in Fig.~\ref{fig:potential_3d}(a), 
where we plot the $\eta$-nucleus optical
potentials as functions of radial coordinate $r$ at $\eta$ energies.
For instance, 
at the $\eta$ threshold $\omega=m_\eta$, the optical potential
has a repulsive core inside the 
nucleus and an attractive pocket on the surface.
Furthermore, this mass reduction yields also the strong energy
dependence of the $\eta$-nucleus optical potential as shown in
Fig.~\ref{fig:potential_3d}(a).
At $\omega\simeq m_\eta-100$ MeV, the real part of the optical potential
is about $140$ MeV attractive while that of the $\eta$ threshold
$\omega=m_\eta$ is $55$ MeV repulsive at $r=0$.

In order to discuss the experimental feasibilities of mesic-nuclei
formations, it is very important to estimate the imaginary parts of the
optical potentials.
As for the $N^*$ width in the medium which is the source of the
imaginary potentials in the present model, we consider the two dominant decay
channels of $N^*$ as $N^*\rightarrow N\pi$ and $NN^*\rightarrow
NN\pi$~\cite{PRC66(02)045202,PRC68(03)035205}.
The detailed evaluation is given in Ref.~\cite{levelcross}.

In the chiral doublet model, there are two possible models concerning
the assignment of the axial charge: the naive and mirror assignments~\cite{Jido:2001nt,Jido:1998av}. 
In this paper, we show only the results with the naive assignment of the
chiral doublet model.
The difference between two assignments in this study appears only in the
estimation of the $N^*$ width in nuclear medium, because the
representation of the mass gap $m^*_N(\rho)-m^*_{N^*}(\rho)$
is same in
the both assignments.
We have already checked the calculated spectra with the mirror
assignment are almost same as that of the naive assignment~\cite{PRC68(03)035205,Nagahiro:2005gf}.

In the second approach to description of $N^*$, which is the  chiral unitary
model, 
the optical potential has quite different features from the previous case. 
In this model, it was found that $N^*$ has a dominant component of 
the $K\Sigma$ channel~\cite{Kaiser:1995cy,Inoue:2001ip,Kolomeitsev:2003kt}.
Since the $\Sigma$ hyperon is free from the 
Pauli blocking in the nuclear medium, only tiny change of the mass gap  is 
expected in the nuclear medium~\cite{Waas:1997pe,Inoue:2002xw}. 
Therefore, the $\eta$ optical potential is attractive
in the bound energy region, $\omega \le m_\eta$, as shown in 
Fig.~\ref{fig:potential_3d}(b), and has weaker energy dependence
compared with that of the chiral doublet model.
We evaluate the $\eta$ optical potential in the
chiral unitary model using the $\eta$ self-energy obtained in
Ref.~\cite{Inoue:2002xw}.
The binding energies and widths of the $\eta$ bound states obtained
with the chiral unitary model are reported in Ref.~\cite{PLB550}.

In recent works~\cite{Hyodo:2008xr}, it was pointed out
that the $N^*(1535)$ obtained in the chiral unitary model
could have some components other
than the state generated dynamically by meson-baryon scattering,
such as genuine quark states,
and that these components could be sources of the chiral
partner of the $N^*(1535)$.
The would-be quark components are implemented in
the subtraction constants, which are model parameters in the
chiral unitary approach. In the present model for $N^*(1535)$
in nuclear medium, the subtraction constants were assumed
to be fixed. This means that the genuine quark components
are independent on medium modifications in this model.
This could be the origin of the different predictions on the in-medium
$N^*(1535)$ mass.

\section{($\pi,N$) reaction for the formation of the $\eta$-nucleus
system} 
\label{sec:formulation}
\subsection{Formulation}
In the beginning of this section, we give the formulation to calculate the
formation spectra of the 
$\eta$-mesic nuclei by ($\pi^+$,p) reaction. We use the same formulation
used in Ref.~\cite{Nagahiro:2005gf}, in which the ($\gamma,$p) reaction 
was discussed for the $\eta$-mesic nuclei. 

To evaluate  the
formation cross section, we use the Green's function method~\cite{Green}.
In this method, the reaction cross section is assumed to be
separated into the nuclear response function $R(E)$ and the elementary
cross section of the $\pi^+n\rightarrow p\eta$ process with the impulse
approximation:
\begin{equation}
\left(\frac{d^2\sigma}{d\Omega dE}\right)
_{A(\pi^+,p)(A-1)\otimes\eta}
=\left(\frac{d\sigma}{d\Omega}\right)
^{lab}_{n(\pi^+,p)\eta}\times R(E),
\end{equation}
where the nuclear response function $R(E)$ is given in terms of the in-medium 
Green's function $G(E)$ as 
\begin{equation}
  R(E) = -\frac{1}{\pi} {\rm Im} \sum_{f} {\cal T}_f^{\dagger} G(E) {\cal T}_f
\end{equation}
where the summation is taken inclusively over all possible final states. The
amplitude ${\cal T}_f$ describes the transition of the incident $\pi$ to
a neutron hole and the outgoing proton: 
\begin{equation}
  {\cal T}_f({\bf r}) = \chi^{*}_{p}({\bf r}) \xi^{*}_{1/2,m_{s}} 
  \left[Y_{l_{\eta}}^{*}(\hat r)\otimes \psi_{j_{n}}({\bf r})\right]_{JM} \chi_{\pi}({\bf r})
\label{eq:ampT}
\end{equation} 
with the neutron hole wavefunction $\psi_{j_n}$, 
the distorted waves of $\pi$ and the ejected proton $\chi_{\pi}$ and $\chi_{p}$,  
the $\eta$ angular wavefunction $Y_{l_{\eta}}(\hat r)$ and the
spin wavefunction $\xi_{1/2,m_s}$ of the ejected proton.
For the neutron hole, we use the harmonic oscillator wavefunction.
The Green's function $G(E)$ contains 
the $\eta$-nucleus optical potential in the Hamiltonian as 
\begin{equation}
   G(E; {\bf r}, {\bf r^{\prime}}) = \langle n^{-1} | \phi_{\eta}({\bf r})
\frac{1}{E-H_{\eta}+i\epsilon} \phi_{\eta}^{\dagger}({\bf r^{\prime}}) |
n^{-1} \rangle
\label{eq:Green_function}
\end{equation}
where $\phi^{\dagger}_{\eta}$ is the $\eta$ creation operator and
$|n^{-1}\rangle$ is the neutron hole state. 

As for the evaluation of the distortion effect of the ($\pi^+$,p)
reaction, we use the eikonal approximation for the description of the
distorted waves of the incoming pion $\chi_\pi$ and of the outgoing
proton $\chi_p$ as,
\begin{equation}
\chi_p^*({\bf r})\chi_\pi({\bf r}) = \exp\left[i{\bf q}\cdot{\bf r}\right]
F({\bf r})
\end{equation}
with the momentum transfer between pion and proton
${\bf q}={\bf p}_\pi-{\bf p}_p$. 
The distortion factor $F({\bf r})$
is defined as in Eqs.~(17) and (18) in Ref.~\cite{Nagahiro:2005gf},
using pion-nucleon and proton-nucleon total cross
sections~\cite{Amsler:2008zz} to take into account the distortion
effects to projectile(pion) and ejectile(proton). 

The calculation of the formation spectra is done separately in 
subcomponents of the $\eta$-mesic nuclei label by $(n\ell_{j})^{-1}_{n}\otimes
\ell_{\eta}$, which means a configuration of a neutron-hole
in the $\ell$ orbit  with the total spin $j$ and the principle quantum 
number $n$ in the daughter nucleus
and an $\eta$ meson in the $\ell_{\eta}$ orbit.
The total formation spectra are obtained by summing up the 
subcomponents, in which we take account of the separation 
energies to make the neutron holes by shifting  the energy relatively.

\subsection{Incident pion energy and elementary cross section}

We choose the incident pion energies so as to satisfy the recoil
free kinematics for the $\eta$ meson production in nuclei. 
In this kinematics, $\eta$ mesons can be created almost at rest
in nuclei. The good advantage is that the formation spectra 
have less subcomponents due to the angular momentum selection 
rule for the $\eta$ and neutron hole states~\cite{EPJA6,PRC66(02)045202,PRC68(03)035205,Nagahiro:2005gf,levelcross}, 
which  
makes the interpretation of the spectra much easier. 

\begin{figure}[hbt]
\center
\resizebox{0.35\textwidth}{!}{%
  \includegraphics{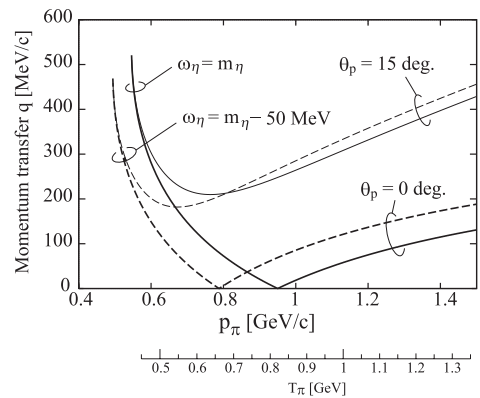}
}
\caption{Momentum transfers at $\eta$ meson energies
 $\omega_\eta=m_\eta$ and $m_\eta-50$ MeV
as
 functions of the incident pion momentum $p_\pi$.
$\theta_p$ denotes the emitted proton angle in the laboratory frame.
The corresponding scale of the pion kinetic energy $T_\pi$ is also
 shown. 
}
\label{fig:mom_trans}       
\end{figure}

In Fig.~\ref{fig:mom_trans}, we plot the momentum transfer of
the A($\pi^+$,p)(A$-1$)$_{\eta}$ reaction for the $\eta$ energies
$\omega=m_{\eta},\ m_{\eta}-50$ MeV with the proton angle 
being 0 and 15 degrees as functions of the incident
pion momentum and pion kinetic energy. In the calculation of 
the momentum transfer, we take heavy mass limit for 
the initial and final nuclei.
As Fig.~\ref{fig:mom_trans} shows, depending on the $\eta$ binding 
energy,  this reaction with $\theta_{p}=0$ 
has the magic momenta for the incident pion 
where the recoilless condition is satisfied. 
In this paper, we set the incident pion kinetic
energies to be 
$T_\pi=820$ MeV and $T_\pi=650$ MeV to satisfy the recoilless condition
at
$\eta$ threshold and $\omega_\eta=m_\eta - 50$ MeV, respectively.
We stress here that 
these energies of the pion beam will be available at J-PARC facility~\cite{Kenta}.

We estimate the elementary cross section
$\left(\frac{d\sigma}{d\Omega}\right)_{n(\pi^+,p)\eta}^{lab}$ using the
experimental data of $\pi^-p\rightarrow n \eta$ measured by Crystal Ball
collaboration~\cite{Prakhov:2005qb}. 
We make use of isospin symmetry to obtain the cross section 
of $\pi^+n\rightarrow p \eta$ from that of the Crystal Ball data. 
We calculate the $\eta$-mesic nucleus formation spectra with 
the incident pion energies of $T_{\pi} = 650,\ 820$ MeV in the laboratory 
frame. 
For these energies, we use the following values of the elementary cross 
section of  $\pi^+n\rightarrow p \eta$:
$
\left(\frac{d\sigma}{d\Omega}
\right)_{n(\pi^+,p)\eta}^{\rm lab} = 2.4$ mb/sr
 for $T_\pi=650$ MeV and 
$
\left(\frac{d\sigma}{d\Omega}
\right)_{n(\pi^+,p)\eta}^{\rm lab} = 0.64$ mb/sr
for $T_\pi=820$ MeV. The former value is read from the existent 
experimental data, whereas the latter is taken from the partial-wave 
analysis (PAW) labeled by I375~\cite{Prakhov:2005qb}, which is almost 
equivalent to the PAW FA02~\cite{Arndt:2003if}.

\subsection{Numerical results of the inclusive ($\pi,N$) spectra}
\label{sec:results}
\begin{figure*}[t]
\center
\resizebox{0.9\textwidth}{!}{%
\subfigure[$T_\pi=820$ MeV]{
\includegraphics{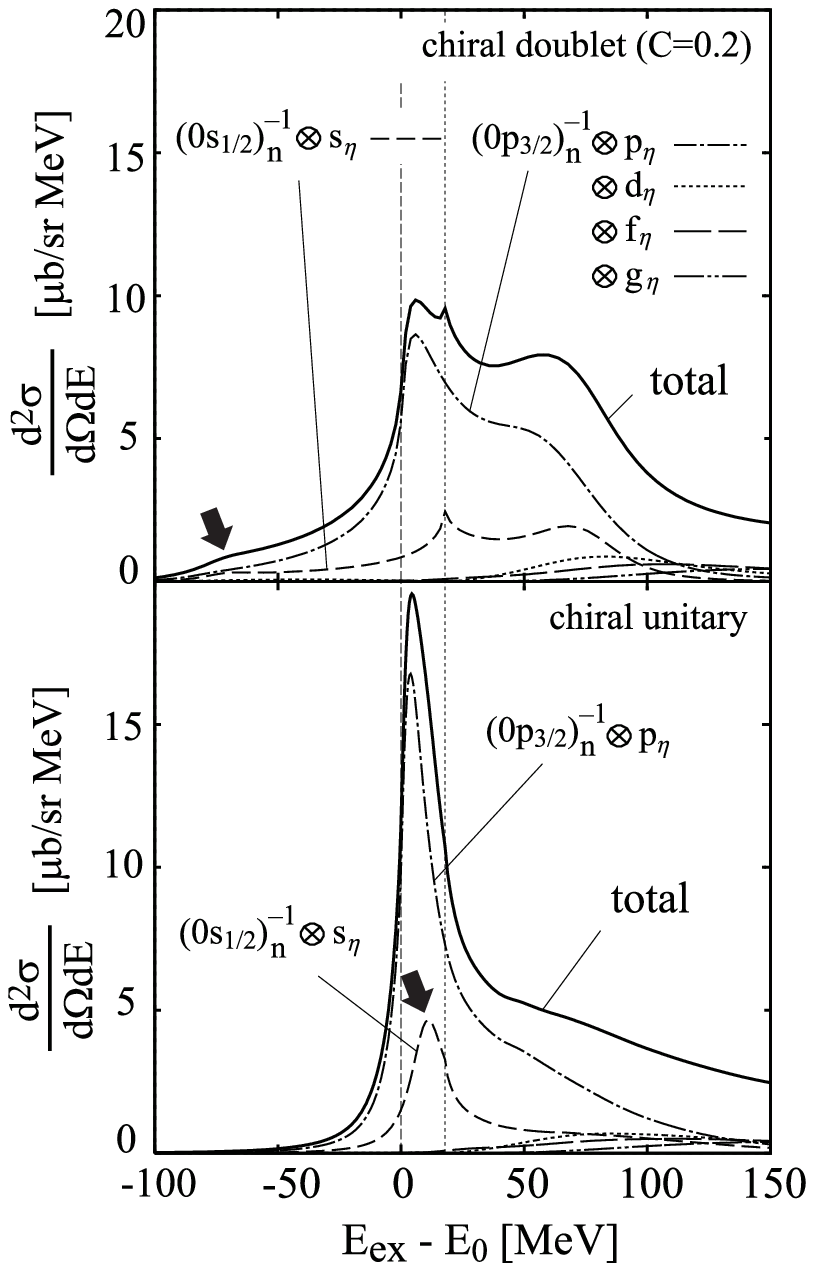}
\label{fig:820}       
}
\hspace{0.1\textwidth}
\subfigure[$T_\pi=650$ MeV]{
\includegraphics{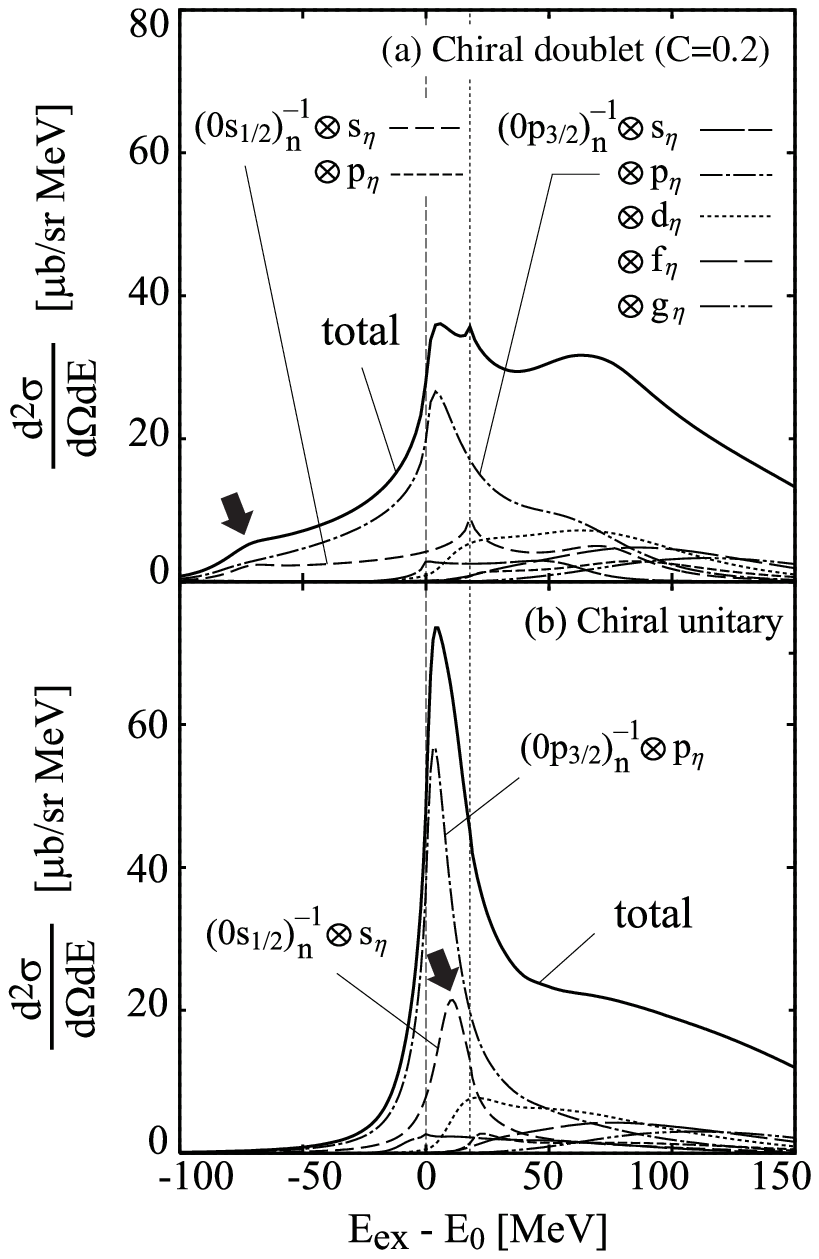}
\label{fig:650}       
}
}
\caption{
Calculated spectra of $^{12}$C($\pi^+$,p)$^{11}$C$\otimes\eta$ reaction
 at (1) $T_\pi=820$ MeV and (2) $T_\pi=650$ MeV and the proton angle $\theta_p=0^\circ$
as functions of the excited energy $E_{\rm ex}$.
$E_0$ is the $\eta$ production threshold. The $\eta$-nucleus
 interaction is calculated by using the chiral doublet model with
 $C=0.2$ (upper panels)
and the chiral unitary model (lower panels).
The thick solid lines show the total spectra and dashed lines
 represent dominant subcomponents as indicated in the figures.
The neutron-hole states are indicated as $(n\ell_j)_n^{-1}$
and the $\eta$ states as $\ell_\eta$.
Solid arrow indicates the peak corresponding to the bound state in each model.
\label{fig:820_650}
}
\end{figure*}

In Fig.~\ref{fig:820}, we show the
$^{12}$C($\pi^+$,p)$^{11}$C$\otimes\eta$ cross sections for the formation
of the $\eta$-$^{11}$C system
in the chiral doublet model with $C=0.2$
(upper panel in Fig.~\ref{fig:820}) and the chiral unitary model
(lower panel in Fig.~\ref{fig:820}).
The incident pion kinetic energy $T_\pi$ is $820$ MeV corresponding to
the recoilless at the $\eta$ threshold. The horizontal axis indicates
the excitation energy $E_{\rm ex}$ measured by the $\eta$ production
threshold $E_0$.
In the figure, we show the total spectra in solid line and the 
contributions from several subcomponents in dashed lines, separately. 
For the spectra of the subcomponents with the $(0s_{1/2})_n^{-1}$ 
neutron-hole state, which is an excited state of the daughter nucleus,
the separation energy 18 MeV is taken into account. Thus the $\eta$
meson production threshold appears at $E_{\rm
ex}-E_0=18$~MeV as indicated in Fig.~\ref{fig:820_650} by the vertical 
dotted line.
The Fig.~\ref{fig:820} shows that the spectra are dominated by two contributions,
$(0s_{1/2})_n^{-1}\otimes s_\eta$ and
$(0p_{3/2})_n^{-1}\otimes p_\eta$, since the final states
with the total spin $J\sim 0$ are largely enhanced under the recoilless
kinematics.
%

%

Let us see the spectra around the threshold; $-50$ MeV $\lesssim
E_{\rm ex}-E_0 \lesssim 50$ MeV.
The spectra in this energy region were already shown in the case
of the (d,$^3$He) and ($\gamma$,p) reactions in
Refs.~\cite{PRC66(02)045202,PRC68(03)035205,Nagahiro:2005gf}. 
The present work confirms that the spectrum shape are very similar
with the previous calculations, showing that the structure of the formation 
spectra is not sensitive to  the reaction mechanism. As already discussed 
in detail in Refs.~\cite{PRC66(02)045202,PRC68(03)035205,Nagahiro:2005gf},
the spectra obtained also in the ($\pi^+$,p) reaction around the $\eta$ production 
threshold show that the repulsive nature of the 
optical potential in the chiral doublet model shifts the spectra into
the higher energy region, whereas the spectra obtained in the chiral unitary model
is shifted into the lower energy region as a result of its attractive
potential. 

Even with the distortion effects to the incident pion, we find that the
difference of expected spectra in the ($\pi^+$,p) reaction between two
approaches for $N^*$ seems to be visible as the case with the
$(\gamma$,p) reaction where the incident $\gamma$ has no distortion effects.

In the case with the chiral doublet model, we can see the
cusp structure in the 
$(0s_{1/2})_n^{-1}\otimes s_\eta$ subcomponent at $E_{\rm
ex}-E_0=18$~MeV corresponding to the $\eta$ threshold for
$(0s_{1/2})_n^{-1}$ hole states.
This is so-called $s$-wave resonance which is found in the
case with weak attraction. In this case with the chiral doublet model,
the surface attractive pocket in the optical potential at the threshold
plays the role of the weak attraction.
In other words, 
we might be able to know the evidence of the curious
shape of the optical potential in the doublet model from the spectral
broadening into the higher energy region (associated with the repulsion) 
together with the cusp structure at the threshold (associated with the
weak attraction), if we could observe it.
However, we should mention here that the
$s$-hole state, corresponding to the excited state of the
daughter nucleus, has natural width which is not taken into account in
the present calculations. By considering the width of the $s$-hole
state, the cusp structure could be smeared out from the spectrum.

Next, let us discuss the bound state structures.
As reported in Refs.~\cite{PRC68(03)035205,Nagahiro:2005gf}, in chiral
unitary model, we can see 
the bound state peak in the subcomponent $(0s_{1/2})_n^{-1}\otimes s_\eta$
around $E_{\rm ex}-E_0\sim 10$ -- $15$ MeV ($\omega_\eta-m_\eta\sim -10$
-- $-5$ MeV) 
as indicated in the lower panel of Fig.~\ref{fig:820}.
The existence of the bound state in the unitary model is predicted in
Ref.~\cite{PLB550}. As seen in the figure, however, it is impossible 
to observe the signals of the bound state, because there is large 
contribution from the $(0p_{2/3})_n^{-1}\otimes p_\eta$ subcomponent
in the same energy and it masks the bound state peak in 
the $(0s_{1/2})_n^{-1}\otimes s_\eta$ component.

For the bound states in the case of the chiral doublet model, as reported 
in Ref.~\cite{levelcross}, there are some bound states obtained 
as solutions in the complex energy plane 
of the Klein-Gordon equation with  the optical 
potential~(\ref{eq:potential}) in the chiral doublet model with 
$C=0.2$\footnote{In Ref.~\cite{PRC66(02)045202}, 
we did not search bound states with the chiral doublet potential $C=0.2$ in 
energies $\omega-m_\eta \lesssim -50$ MeV, because such deep energies 
were out of scope of investigation in Ref.~\cite{PRC66(02)045202}.}.
For $\eta$ with $\ell_{\eta}=0$, the bound states were found
with the eigenenergies (B.E., $\Gamma$)=(91.3, 26.3) MeV for $0s$
and (75.1, 33.0) MeV for $1s$~\cite{levelcross}.
In Fig.~\ref{fig:820_650} we can see that
the $0s$ bound state appears 
in the spectrum as a bump around $E_{\rm ex}-E_0\sim -70$ -- $-80$
MeV. 
These deep bound states correspond to the lower mode mentioned in
Sec.~\ref{sec:level_cross} in the chiral doublet model ($C=0.2$).
However, the strength of the bump in the spectrum is too small 
to be observed in experiments.
The $\eta$ bound states with $\ell_{\eta}=1$ were also obtained 
at (79.3, 31.1) MeV for $0p$ and
(72.1, 34.2) MeV for $1p$. The corresponding peak structure 
to the $0p$ bound state is not seen in the spectrum again due to 
the small strength. 
In the recoilless kinematics, the $1s$ and $1p$ bound states should 
give much less contributions in the spectrum, 
since there are no $1s$- and $1p$-hole
states in the daughter nucleus for the carbon target case.

Much more prominent structure is seen in the quasi-free region
$E_{\rm ex}-E_0>0$ in the case of the chiral doublet model.  
As also discussed in detail in Ref.~\cite{levelcross}, we see considerably
large 
bump structure around $E_{\rm ex}-E_0\sim 60$ MeV as shown in
Fig.~\ref{fig:820}(upper panel). 
This peak
comes from the $N^*$-{\it hole} mode coupled to the $\eta$ meson in the
medium, namely the upper mode shown in Fig.~\ref{fig:map}(b).
As discussed in Sec.~\ref{sec:level_cross}, the upper mode in the chiral
doublet model is enhanced as a consequence of the level crossing
associated with the reduction of the $N$-$N^{*}$ mass gap in nuclear
matter stemming from the partial restoration of the chiral symmetry.
On the other hand, in the case of the chiral unitary approach, we 
see smooth slope in the quasi-free region. Hence, the enhancement
in the quasi-free region could be a signal for the reduction of 
the $N$-$N^*$ mass gap in nuclear medium, which supports 
the scenario of partial restoration of  chiral symmetry in nuclear medium.

Finally, we mention the incident pion energy dependence of the spectra.
In Fig.~\ref{fig:650}, we show the spectra with $T_\pi=650$ MeV
corresponding to the recoilless at $\omega_\eta-m_\eta\sim -50$ MeV.
By setting the recoilless condition here, we have expected some
enhancement of the deep bound state in the 
chiral doublet model.
However, although small enhancement of the bound state can be seen, its
effect is not large enough to make this bump clearly observed.

In $T_\pi=650$ MeV case, we find that the magnitude of the cross section
is four times larger than that of $T_\pi=820$ MeV, because of the larger
elementary cross section. We also find that, due to
the relatively large momentum transfer ($\sim 100$ MeV/c) at $T_\pi=650$ MeV, many
subcomponents give finite contributions at quasi-free region, while in
the $T_\pi=820$ MeV case only two subcomponent $(0p_{3/2})_n^{-1}\otimes
p_\eta$ and $(0s_{1/2})_n^{-1}\otimes s_\eta$ at almost all energy
region shown there.

\subsection{Spectra for $\pi N$ coincident observation}

\begin{figure}[htb]
\center
\resizebox{0.35\textwidth}{!}{%
  \includegraphics{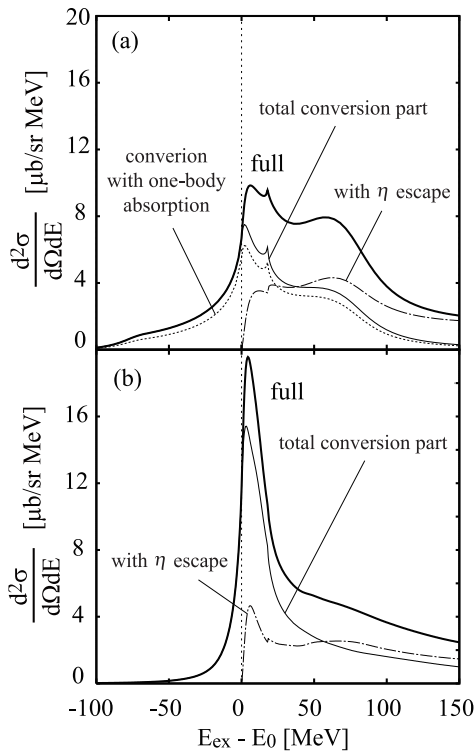}
}
\vspace{5mm}       
\caption{Decomposition of the full spectra into the conversion parts and
 the escape part which are defined in 
 the text. The reaction and energy are $^{12}$C($\pi^+$,p)$^{11}$C$\otimes\eta$ and
 $T_\pi=650$ MeV in (a) with the chiral doublet model and (b) the chiral
 unitary model.
The full spectra are shown by the thick solid line and the
 conversion parts, which includes both of decay channels of
 $N^*\rightarrow N\pi$ and $N^*N\rightarrow NN\pi$, is shown by the thin
 solid line. The dashed line denotes the 
spectrum including only the $N^*\rightarrow N\pi$
 conversion part, and the dot-dashed line denotes the escape part.
}
\label{fig:Scon}       
\end{figure}

In inclusive measurements, it may be hard to separate signals out
of large backgrounds. It was argued
for the previous ($\pi^{+}$,p) experiments in finite momentum 
transfer in Refs.~\cite{Chrien:1988gn,Haider}
that possible backgrounds could come from quasi-free knockout, 
multiple pion and proton scattering, and pion absorption and that 
the signal-noise ratio was estimated  to be 1/10.
It would be also expected in the present setup for the ($\pi^+$,p)
reaction that the background can be as large as the previous
experiment. 
To subtract such large background, it is useful to take some
coincidence measurements accompanying the $\eta$ meson production 
in nuclei, for example simultaneous observation of $N\pi$ pair coming from 
$N^*$ decay in a nucleus~\cite{sokol,Jha}.

In the Green's function method~\cite{Green}, one can 
separately calculate each contribution to the spectrum coming from
the different $\eta$ absorption 
process. Along to the formalism in Ref.~\cite{Green}, we rewrite
equivalently the
imaginary part of the Green's function of $\eta$ as
\begin{equation}
{\rm Im}G = (1+G^\dagger V_\eta^\dagger){\rm Im}G_0 (1+V_\eta G) + G^\dagger{\rm Im}V_\eta G 
\label{eq:Green}
\end{equation}
where $G$ and $G_0$ denote the full and free Green's functions for $\eta$
and $V_\eta$ is the $\eta$ optical potential.
The first term of the right-hand-side of
Eq.~(\ref{eq:Green}) represents the contributions from the escape of
$\eta$ from daughter nuclei and the second term describes the
conversion process caused by the $\eta$ absorption into the nuclei. 
Thus, calculating only the conversion part, we can show spectra associated
with decays (or absorptions) of $\eta$ mesons in nuclei, which can be obtained
by observing emitted particles. 

In the chiral doublet model, we take into account two
decay channels of $N^*$ in nuclear medium as mentioned in
Sec.~\ref{sec:potential}; $N^*\rightarrow N\pi$ and $N^*N\rightarrow
NN\pi$. These contributions are expressed in the conversion part. 
Therefore, we can separate the spectra further into two terms by describing 
the conversion part as 
\begin{equation}
{\rm Im}V_\eta={\rm Im}V_{\eta}(N^*\rightarrow N\pi)  +{\rm Im}V_{\eta}(N^*N\rightarrow NN\pi).
\label{eq:ImU}
\end{equation}

As shown in Fig.~\ref{fig:Scon}(a), we decompose the total spectra shown
in Fig.~\ref{fig:820}(upper panel) into three parts; the contribution from the
$\eta$-escape process, the conversion part of the $N^*\rightarrow N\pi$ and
that of $N^*N\rightarrow NN\pi$ processes. The expected spectra with the
coincidence of the $N\pi$ pair from $N^*$ is indicated by the dotted
line in the figure. 
The thin solid line in Fig.~\ref{fig:Scon}(a) includes both of the
one-body 
and two-body absorption of $N^*$ in the chiral doublet model.
It is found that the $N^*N\rightarrow NN\pi$ contribution is much
smaller than that of $N^*\rightarrow\pi N$.
We also find that, in the doublet model case, the strength of the peak
structure in the quasi-free region corresponding to the $N^*$-{\it hole}
mode is reduced to be about half by taking the coincidence of $N\pi$
pair from $N^*$.

In Fig.~\ref{fig:Scon}(b), we plot the same figure for the chiral unitary 
model as Fig.~\ref{fig:Scon}(a). We only show the total conversion part,
since decomposition of the $\eta$ self-energy into different absorption 
processes was not done in the chiral unitary model. 
We expect that the difference between two approaches with
and without the $N^*$ mass reduction is 
still large even we take the coincidence of the $N^*$ decay.
In these estimation we do not take into account any final state interaction 
for emitted particles from $N^*$ decay in a nucleus.
These contributions could be important for further
qualitative discussions.

For the background estimation in experiments,
it would be useful to observe the ($\pi^-$,p) process in the same setup
as the ($\pi^+$,p). In the ($\pi^-$,p) process $\eta$ mesons cannot be
created due to isospin symmetry, and it is expected that the formation
spectra have no structure at the $\eta$ meson threshold. This means that
all the contributions are regarded as ``background''.

\section{Additional discussions on experimental feasibility}
\label{sec:feasability}
\subsection{Comparison with the ($\pi^+$,p) reaction measured at Brookhaven}
The
($\pi^+$,p) reaction experiment 
for the formation of the $\eta$-mesic nuclei
has been already performed at Brookhaven in
1988~\cite{Chrien:1988gn}.
The experiment have been done at a finite proton angle $\theta_p=15^\circ$
in the laboratory frame, based on the theoretical suggestion~\cite{Haider}
to observe narrow peaks.  
The recoilless condition cannot be satisfied
for finite angle proton emissions. In the case of  $\theta_p=15^\circ$,
the momentum transfer is  larger than 200
MeV/c in any incident pion momentum as shown in Fig.~\ref{fig:mom_trans}.
Therefore, the expected spectra will be completely different from that
of the forward angle.

\begin{figure}[hbt]
\center
\resizebox{0.4\textwidth}{!}{%
  \includegraphics{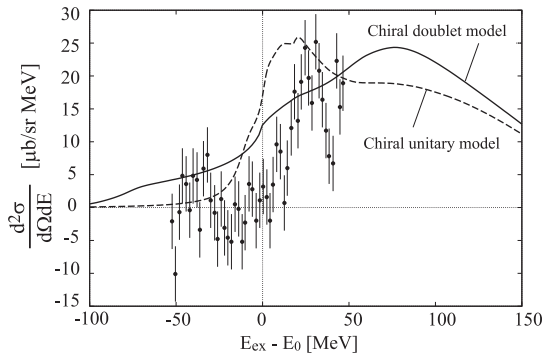}
}
\caption{
Comparison of the calculated total spectra with the signal part of the
 experimental data on the carbon target case reported in
 Ref.~\cite{Chrien:1988gn} after the background subtraction shown in the
 same reference.
The solid line indicates the total spectrum with the chiral doublet
 model ($C=0.2$) and the dashed line is that of the chiral unitary
 model, calculated at the same condition with the data.
In the theoretical calculations, the angular momentum of $\eta$ is taken
 into account up to $\ell_\eta=6$. 
}
\label{fig:Chrien}       
\end{figure}

We calculate the ($\pi^+$,p) spectra with the proton angle 
$\theta_p=15^\circ$ and the incident pion energy 
$T_\pi=673$ MeV ($p_\pi=800$ MeV/c) in the same theoretical 
procedure as in the previous section, in order to compare the 
theoretical calculations with the experimental data obtained at
Brookhaven. The comparison for the carbon target case is shown 
in Fig.~\ref{fig:Chrien}. The data are taken from  the second figure of
Fig.~1 in Ref.~\cite{Chrien:1988gn} after subtraction of background 
estimated in the paper. They showed an experimental error bar only for
one experimental point.  We assume that the single error bar is a 
typical error for all the point and put the same error bar to all the point. 
We find in Fig.~\ref{fig:Chrien} 
that both models provide the consistent results with the
experimental data. This means that the experiment with the finite proton
angle in Ref.~\cite{Chrien:1988gn}
is not sensitive to the in-medium
properties of $N^*$~\cite{PLB231,Kohno:1990xv}.

\begin{figure}[th]
\center
\resizebox{0.4\textwidth}{!}{%
  \includegraphics{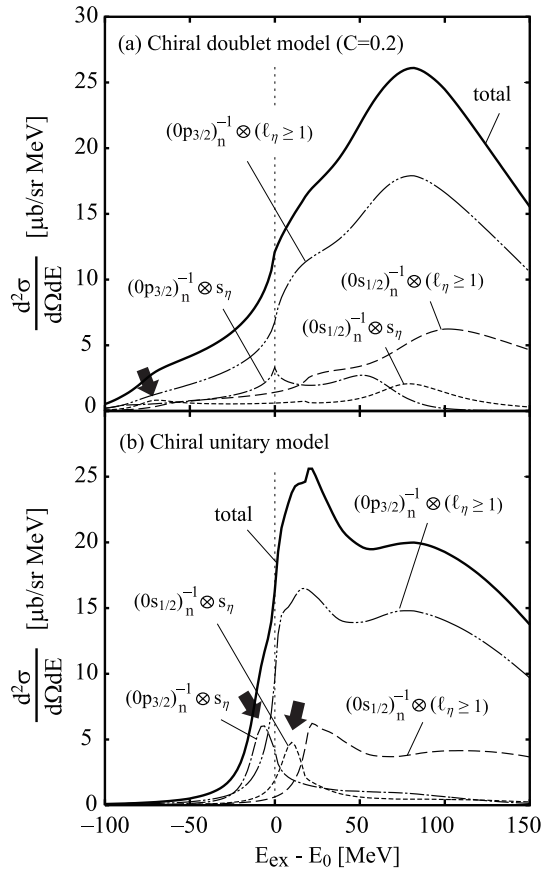}
}
\caption{Calculated spectra of $^{12}$C($\pi^+$,p)$^{11}$C$\otimes\eta$ reaction
 at $T_\pi=673$ MeV ($p_\pi=800$ MeV/c) and the proton angle $\theta_p=15^\circ$
as functions of the excited energy $E_{\rm ex}$.
$E_0$ is the $\eta$ production threshold. The $\eta$-nucleus
 interaction is calculated by (a) the chiral doublet model with $C=0.2$
and (b) the chiral unitary model.
The thick solid lines show the total spectra and each dashed line
 represents the dominant subcomponent as indicated in the figures.
The neutron-hole states are indicated as $(n\ell_j)_n^{-1}$
and the $\eta$ states as $\ell_\eta$.
The angular momentum of $\eta$ is taken
 into account up to $\ell_\eta=6$.
Solid arrows indicate the peaks corresponding to the bound states.
}
\label{fig:15deg}       
\end{figure}

We show the details of the results calculated theoretically with finite proton angle in
Fig.~\ref{fig:15deg}.
We calculate the spectra by taking into account the $\eta$ angular momenta
up to $\ell_{\eta}=6$. In the case of the finite proton angle, in contrast
to the forward proton, many subcomponents have substantial contributions to the
total spectra, especially higher angular momentum components $\ell_\eta \ge 1$, 
because the strong suppression for $J\neq 0$ configuration is not made 
any more. This makes it difficult to interpret the spectrum structure 
in terms of the properties of the $\eta$ and $N^{*}$ in nuclear medium. 

In the spectrum calculated with the chiral unitary model, the lower panel
of Fig.~\ref{fig:15deg}, two prominent peaks are seen around 
$E_{ex}-E_{0}=-10$ MeV in the subcomponents 
$(0s_{1/2})^{-1}_n\otimes s_\eta$
and $(0p_{3/2})^{-1}_n\otimes s_\eta$\footnote{In the forward proton case,
this subcomponent $(0p_{3/2})^{-1}_n\otimes s_\eta$ is strongly suppressed 
in the recoilless kinematics.}. These are bound state
signatures of the $\eta$ meson with $\ell_{\eta}=0$ in nuclei. In particular,
it is interesting that 
the peak of $(0p_{3/2})^{-1}_n\otimes s_\eta$ appears in bound region,
where contamination from the quasi-free $\eta$ contributions could be
expected to be small. 
In fact,
this observation was the original idea by Haider and Liu for the advantage 
of the finite proton angle experiment. As seen in the figure, however, 
this bound state peak
is masked by the tail of  the quasi-free contribution of
$(0p_{3/2})^{-1}_n\otimes (\ell_\eta \ge 1)$ caused by 
virtual $\eta$ absorption in the large imaginary potential. Thus, even 
if the chiral unitary model prediction of the bound state is correct,
the bound state signal cannot be observed separately from the quasi-free
contribution in the total spectrum.
The same situation might have occurred in the Brookhaven
experiment~\cite{Chrien:1988gn}. 
Therefore, the attempt to separate a shallow bound state from the
quasi-free 
contribution by setting $\theta_p=15^\circ$
seems not to work when the imaginary potential is large.

By all considerations mentioned above, we think
that it is better to set the final proton angle
to be zero degree where the difference reflecting 
the distinct in-medium $N^*$ properties 
expected to be larger.
The energy range measured by the experiment~\cite{Chrien:1988gn}
did not cover whole energies of the emitted proton where 
the interesting features, like the deep bound state and the bump structure in the
quasi-free region caused by possible level crossing phenomena, take place.

\subsection{$(\pi^-,n)$ reaction with the $^7$Li target}

Here we show the formation spectra of the $\eta$-mesic nuclei 
in the $(\pi^-,n)$ reaction with the $^{7}$Li target. 
As shown in Figs.~\ref{fig:820} and \ref{fig:650} for the carbon target, 
the total spectra were
dominated by the $p$-wave components $(0p_{2/3})_n^{-1}\otimes p_\eta$
since the target nucleus $^{12}$C has four neutrons in $p$-state while
two in $s$-state.
Then, the $s$-wave components are
relatively small, which contain the interesting structure such as the
bound states and/or the threshold cusp.

\begin{figure}
\center
\vspace{5mm} 
\resizebox{0.35\textwidth}{!}{%
  \includegraphics{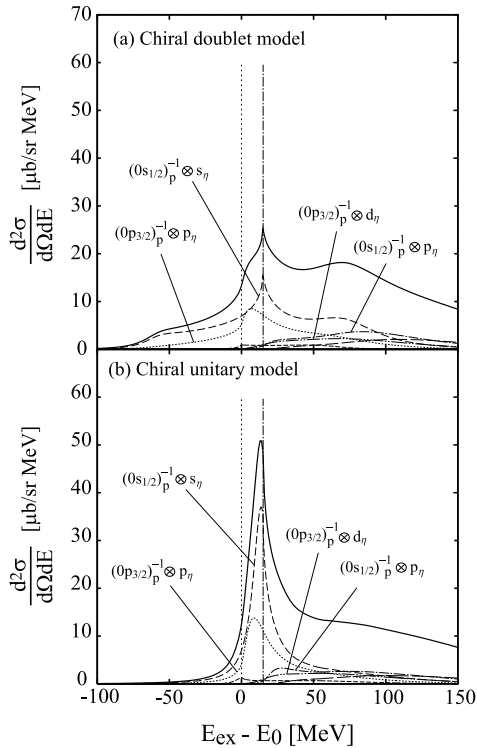}
}
\caption{Calculated spectra of $^{7}$Li($\pi^-$,n)$^{6}$He$\otimes\eta$ reaction
 at $T_\pi=650$ MeV and the proton angle $\theta_p=0^\circ$
as functions of the excited energy $E_{\rm ex}$.
$E_0$ is the $\eta$ production threshold. The $\eta$-nucleus
 interaction is calculated by (a) the chiral doublet model with $C=0.2$
and (b) the chiral unitary model.
The thick solid lines show the total spectra and each dashed line
 represents the dominant subcomponent as indicated in the figures.
Here, the proton-hole states are indicated as $(n\ell_j)_p^{-1}$
and the $\eta$ states as $\ell_\eta$.
}
\label{fig:7Li}       
\end{figure}

In Fig.~\ref{fig:7Li}, we show the spectra of $^7$Li($\pi^-,n$) reaction
with the elementary process $\pi^-p\rightarrow n \eta$ (one
proton picked-up). 
We see that the $s$-wave
component now dominates the spectrum for each case (a) and (b),
since $^7$Li has a single proton in $p$-state.
We can find a deep bound state peak
in $s_\eta$ state around $E_{\rm ex}-E_0\sim -60$ MeV in the chiral
doublet model also in $^7$Li target 
case while in the chiral unitary case we can see only the threshold peak
structure. 
We consider that the experimental data with $^7$Li target would be
useful and be a complement to that of $^{12}$C.

\subsection{Consideration of finite experimental resolution}

We also estimate the effect of the finite experimental energy resolution
on the spectra. For this purpose, 
we fold the calculated spectra with
$T_\pi=650$ MeV, which is shown in Fig.~\ref{fig:650}, as
\begin{equation}
\int f(E')g(E-E') dE',
\label{eq:fold}
\end{equation}
where $f(E')$ represents the calculated spectra, and $g(E)$ expresses
the effect of the finite energy resolution and is given in a gaussian
form as
\begin{equation}
g(E-E')=\frac{1}{a\sqrt{\pi}}\exp\left[
-\left(\frac{E-E'}{a}\right)^2
\right]
\end{equation}
\begin{figure}[bh]
\center
\vspace{5mm}
\resizebox{0.35\textwidth}{!}{%
  \includegraphics{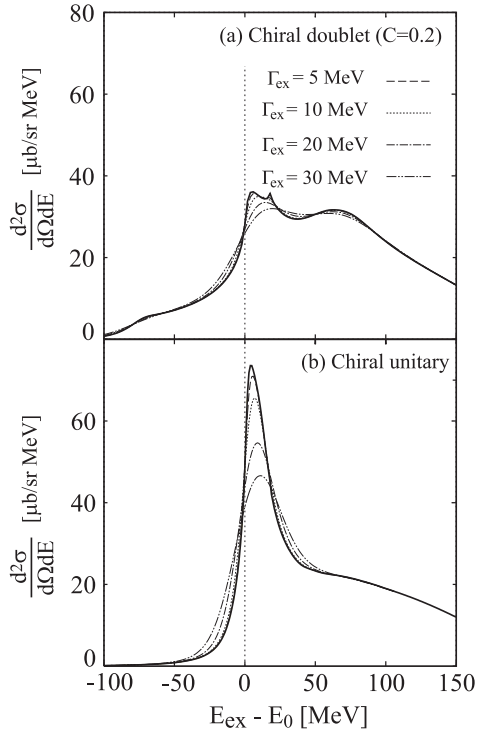}
}
\caption{Convolutions of the spectra of
 $^{12}$C($\pi^+$,p)$^{11}$C$\otimes\eta$ reaction 
 at $T_\pi=650$ MeV and the proton angle $\theta_p=0^\circ$. 
The thik solid lines
 indicate the spectra without convolution, and other lines are spectra
 taken convolutions with finite experimental resolution $\Gamma_{\rm ex}$
 indicated in the figures.}
\label{fig:fold}       
\end{figure}
with $a=\Gamma_{\rm ex}/2\sqrt{\ln 2}$. The 
experimental resolution
(FWHM) is denoted by $\Gamma_{\rm ex}$.
In Fig.~\ref{fig:fold}, we show the calculated result with several energy 
resolutions. 
We can observe the difference of two 
approaches with $\Gamma_{\rm ex}\sim20$ MeV, which are expected to be reached
at J-PARC facility~\cite{Kenta}. 

\section{Conclusion}
\label{sec:conclusion}

We calculated the  $\eta$-mesic nuclei formation spectra
of ($\pi,N$) reactions with nuclear targets
to discuss the experimental feasibility in forthcoming experiments.
Emphasizing in-medium properties of the $N^{*}(1535)$ baryon 
resonance in the context of chiral symmetry,  
we discussed the structure of the $\eta$-mesic nuclei formation 
spectra. Especially, the reduction of $N$-$N^*$ mass gap in the nuclear 
medium were investigated by two chiral models.
In the chiral doublet model, $N^*(1535)$ is regarded as a
chiral partner of nucleon and the  mass gap is expected to be reduced 
in association with the partial restoration of the chiral
symmetry in nuclear medium~\cite{PRC66(02)045202}.
In the chiral unitary approach, $N^*(1535)$ is
introduced as a resonance dynamically generated and described as a
quasi-bound state of the Kaon and Hyperon, and reduction of the mass gap is
expected to be small in nuclear matter~\cite{Waas:1997pe,Inoue:2002xw}.

We showed the formation spectra of $\eta$-mesic nuclei by ($\pi,N$)
reactions. We confirmed that the magnitude of the cross section is large
enough to be observed in experiments.
We found that the ($\pi,N$) reactions were also appropriate to observe
the interesting behaviors like deep bound states and
the bump structure at the quasi-free region which can be understood by
the concept of the level crossing phenomena caused by the partial
restoration of the chiral symmetry~\cite{levelcross}.
We conclude that we can get new information on the in-medium
$N^*$ properties through the formation of the $\eta$-mesic nuclei by
($\pi,N$) reactions.

We also discussed expected background and ways of its reduction by the
simultaneous observation of $N\pi$ pairs from the $N^*$ decay in
medium. We found that the difference between two treatments of in-medium
$N^*$ with and without the reduction of the $N$-$N^{*}$ mass gap 
are not largely affected by the
coincidence observation and there is some chance to make the observation
clearer. 

We believe that the present theoretical results are important to
stimulate both theoretical and, especially, experimental activities to
study the hadron properties in-medium and to obtain new information on
the partial restoration of the chiral symmetry in nuclear medium.

\section*{Acknowledgement}
We would like to express our thanks to E.E.~Kolomeitsev for
fruitful collaboration. We also thank K.~Itahashi and H.~Fujioka for
the useful discussions from experimental side. One of the author (H.~N.)
is the Research Fellow of the Japan
Society for the Promotion of Science (JSPS). This work is partially 
supported
by the Grant for Scientific Research
(No.~18$\cdot$8661, 18042001, 20028004). A part of this work was done
under Yukawa International Project for Quark-Hadron Science (YIPQS).

\end{document}